\documentclass[dvips,11pt]{article}
\usepackage{varioref,exscale,latexsym,amsmath,amssymb}
\usepackage{graphicx}

\def\beq{\begin{equation}}

\def\eeq{\end{equation}}

\def\beqa{\begin{eqnarray}}

\def\eeqa{\end{eqnarray}}

\title{{\bf The fine-tuning problems of particle physics and anthropic mechanisms}}
\author{John F. Donoghue   \\ \\
 Department of Physics\\
University of Massachusetts\\
Amherst, MA  01003, USA\\  \\}
\begin{document}
\begin{titlepage}
\maketitle
\begin{abstract}
Many of the classic problems of particle physics appear in a very different light when viewed
from the perspective of the multiverse. Most importantly the two great ``fine tuning'' problems
that motivate the field are far less serious when one accounts for the required anthropic
constraints which exist in a multiverse. However, the challenge then becomes to construct
a realistic physical theory of the multiverse and test it. I describe some phenomenology
of the quark and lepton masses that may provide a window to the theory that underlies the
multiverse.
\end{abstract}
\vspace{0.2 in}

\end{titlepage}
\section*{Open questions in particle physics}

Each field has a set of questions which are universally viewed as important, and these questions
motivate much of the work in the field. In particle physics, several of these questions
are directly related to experimental problems. Examples include questions such as: Does the Higgs
boson exist and if so what is its mass? What is the nature of the dark matter seen in the Universe? What is the mechanism that generated the net number of baryons in the Universe? For these topics, there is a well posed
problem related to experimental findings or theoretical prediction. These are problems that must be solved if we are to achieve a complete understanding of the fundamental theory.

There are also a different set of questions which have a more aesthetic character. In these cases, it is
not as clear that a resolution is required, yet the problems motivate a search for certain classes of theories.
Examples of these are the three ``naturalness'' or ``fine-tuning'' problems of the standard model - those of the cosmological
constant $\Lambda$, the scale of electroweak symmetry breaking $v$ and the strong CP-violating angle $\theta$. As will be explained more fully below, these parameters are free parameters in the Standard Model that seem to have values 10 to
120 {\it orders of magnitude} smaller than their natural values and smaller than the magnitude of their quantum corrections. Thus their ``bare'' value plus their quantum corrections need to be highly fine-tuned in order to obtain the observed values. Because of the magnitude of this fine-tuning, one suspects that there is a dynamical mechanism
at work that makes the fine tuning natural. This motivates many of the theories of new physics beyond the Standard
Model. A second set of aesthetic problems concerns the parameters of the Standard Model, the coupling constants and masses of the theory. While the Standard Model itself is constructed simply using gauge symmetry, the parameters
themselves seem not to be organized in any symmetric fashion. We would love to uncover the principle that organizes the quark and lepton masses (sometimes referred to as the Flavor Problem), for example, but attempts to do so with symmetries or a dynamical mechanism have been unsuccessful.

These aesthetic questions are very powerful motivations for new physics. For example the case for low energy supersymmetry, or other TeV scale dynamics to be uncovered at the Large Hadron Collider (LHC), is based almost entirely on the fine-tuning problem for the scale of electroweak symmetry breaking. If there is new physics at the TeV scale, then there need not be any fine tuning at all and the electroweak scale is natural. We are all greatly looking forward to the results of the LHC, which will tell us if there in fact is new physics at the TeV scale. However, the aesthetic questions are of a different character from direct experimental ones concerning the existence and mass of the Higgs boson. There does not {\it have} to be a resolution to the aesthetic questions - if there is no dynamical
solution to the fine-tuning of the electroweak scale, it would puzzle us but would not upset anything with the
fundamental theory. We would just have to live with the existence of fine-tuning. However, if the Higgs boson is not found within a given mass range it would falsify the Standard Model.

The idea of a multiverse will be seen to drastically change the way that we perceive the aesthetic problems - those of fine-tuning and of flavor. In a multiverse, the parameters of the theory vary from one domain to another.
This naturally leads to the existence of anthropic constraints - only some of these domains will have
parameters that reasonable allow the existence of life. We can only find ourselves in a domain which satisfies
these anthropic constraints. Remarkably, the anthropic constraints provide plausible ``solutions'' to
two of the most severe fine-tuning problems, those of the cosmological constant and the electroweak scale. Multiverse theories also drastically reformulate some of the other problems - such as the flavor problem. However at the same time, these theories raise a new set of issues for new physics. My purpose in this article is to discuss how
the idea of the multiverse reformulates the problems of particle physics.

It should be noted up front that the Anthropic Principle\cite{anthropic} has had a largely negative reputation in the particle physics community. At some level this is surprising - a community devoted to uncovering the underlying fundamental theory might be expected to be interested in exploring an suggestion as fundamental as the Anthropic Principle. I believe that the problem really lies in the word ``Principle'' more than ``Anthropic''. The connotation of ``Principle'' is that of an underlying theory. This leads to debates of whether such a principle is scientific, i.e. can it be tested. However, ``anthropics'' is not itself the theory, nor even a principle. Rather, in certain physical theories the word applies to constraints that naturally occur within the full theory. However it is the theory itself that need to be tested, and to do this one needs to understand the full theory and pull out its predictions. For theories that lead to a multiverse, anthropic
constraints are unavoidable. As we understand better what types of theory has this multiverse property, the word
anthropic is finding more positive applications in the particle community. This article also tries to describe some of the
ways that ``anthropic'' can be used to positive effect in particle physics.

\section*{The golden lagrangian and its parameters}

The Lagrangian of the Standard Model (plus general relativity) encodes our present understanding of all observed physics except for dark matter\cite{dynamics}. The only unobserved ingredient of the theory is the Higgs boson. The Standard Model is built on the principle of gauge symmetry - that the Lagrangian has a $SU(3)\otimes SU(2)_L\otimes U(1)$ symmetry at each point of spacetime. This, plus renormalizeability, is a very powerful constraint and uniquely defines the structure of the Standard Model, up to a small number of choices such as the number of generations of fermions. General relativity is also defined by a gauge symmetry - local coordinate invariance. The results can be written in compact notation as
\begin{eqnarray}
{\cal L} &=& -\frac14 F^2 + \bar{\psi} i D \psi + \frac12 D_\mu \phi D^\mu \phi \nonumber \\
&+& \bar{\psi} \Gamma \psi \phi + \mu^2 \phi^2 -\lambda \phi^4 -\frac{1}{16\pi G_N}R -\Lambda
\end{eqnarray}
Of course, such a simple form belies a very complex theory and tremendous work is required to understand the predictions of the Standard Model. But the greatest lesson of particle physics of the past generation is that
Nature organizes the universe through a simple set of gauge symmetries.

However, the story is not complete. The simple looking Lagrangian and the story of its symmetry based origin also hide a far less beautiful fact. To {\it really} specify the theory we need not only the Lagrangian above, but also have to give a set of 28 numbers which are the parameters of the theory. These are largely hidden underneath the compact notation of the Lagrangian. Examples include the masses of all the quarks and leptons (including neutrinos), the strength of the three gauge interactions, the weak mixing angles describing the charge current interactions of quarks and of leptons, the overall scale of the weak interaction, the cosmological constant and Newton's gravitational constant. None of these parameters are predicted by the theory. The values that have have been uncovered experimentally do not obey any known symmetry pattern, and the Standard Model provides no principle by which to organize them. After the beauty of the Standard Model Lagrangian, these seemingly random parameters reinforce the feeling that
there is more to be understood.

\section*{Fine tuning}

There are three of the 28 parameters which appear especially puzzling, because their values appear to be unnaturally small. Naturalness and fine-tuning have very specific technical meanings in particle physics. These meanings are
related to, but not identical to, the common usage in non-technical settings. The technical version is tied to the magnitude of quantum corrections. When one calculates the properties of any theory using perturbation theory, quantum mechanical effects give additive corrections to all the parameters of the theory. Perturbation theory describes the various quantities of a theory in a power series in the coupling constants. The calculation involves summing over the effects of all
virtual states that are possible in the theory, including those at high energy. The quantum correction refers to the terms in the series that depend on the coupling constants. The ``bare'' value is the term independent of the coupling constants. The physical measured value is the sum of the bare value and the quantum corrections.

The concept of naturalness is tied to the magnitude of the quantum corrections. If the quantum correction is of the same order as (or smaller than) the measured value, the result is said to be natural. If, on the contrary, the measured value is much smaller than the quantum correction the result is unnatural because the bare value and the quantum correction appear to have an unexpected cancelation to give a result that is much smaller than either component. This is an unnatural fine-tuning.

In fact, the quantum correction is often not precisely defined. The ambiguity can arise due to possible uncertainties of the theory at high energy. Since physics is an experimental science, and we are only gradually uncovering the details of the theory as we probe higher energies, we do not know the high energy limits of our present theory. We expect new particles and interactions to be uncovered as we study higher energies. Since the quantum correction includes effects from high energy, there is an uncertainty about their extent and validity. We understand the theory up to some energy - let us call this energy $E_{\rm max}$ - but beyond this energy new physics may enter. The quantum corrections will typically depend on the scale $E_{\rm max}$. We will see below that in some cases the theory may be said to be natural if one employs low values of $E_{\rm max}$, but becomes unnatural for high values.

The Higgs field in the Standard Model takes a constant value everywhere is spacetime. This is called its vacuum-expectation-value, $v$, abbreviated as vev, which has the magnitude $v=246$~GeV. This is the only dimensionful constant in the electroweak interactions and hence sets the scale for all dimensionful parameters of the electroweak theory. For example, all of the quark and lepton masses are given by dimensionless numbers, $\Gamma_i$ (the Yukawa couplings) times the Higgs vev, $m_i=\Gamma_i v/\sqrt{2}$. However the Higgs vev is one of the parameters which has a problem with naturalness. While the vev depends on many parameters, the problem is well illustrated by its dependence on the Higgs coupling to the top quark. In this case, the quantum correction grows quadratically with $E_{\rm max}$. One finds
\begin{equation}
v^2 =v_0^2 + \frac{3\Gamma_t^2}{4\pi^2\lambda} E_{\rm max}^2
\end{equation}
where $v_0$ is the bare value, $\lambda$ is the self coupling of  the Higgs, and the second term is the quantum correction. Since $v = 246$~GeV and $\Gamma,\lambda \sim 1$, this would be considered natural if $E_{\rm max} \sim 10^3$~GeV, but would be unnatural by 26-32 orders of magnitude if $E_{\rm max} \sim 10^{16}$~GeV (characteristic of grand unified theories uniting the electroweak and strong interactions) or $E_{\rm max} \sim 10^{19}$~GeV (characteristic of the Planck mass which sets the scale for quantum gravity).

If we philosophically reject fine-tuning and require that the Standard Model be technically natural, this requires that $E_{\rm max}$ should be around 1~TeV. For this to be true we need a new theory to enter at this scale that removes the quadratic dependence on $E_{\rm max}$. Such theories do exist - supersymmetry is a favorite example. Thus the argument against fine-tuning becomes a powerful motivator for new physics at the scale of 1~TeV. The Large Hadron Collider has been designed to find this new physics.

An even more extreme violation of naturalness involves the cosmological constant $\Lambda$. Experimentally, this dimensionful quantity is of order $\Lambda \sim (10^{-3} ~{\rm eV})^4$. However, the quantum corrections to this quantity grow like the fourth power of the scale $E_{\rm max}$
\begin{equation}
\Lambda = \Lambda_0 + c E_{\rm max}^4
\end{equation}
with $c$ of order unity. This quantity is unnatural for all particle physics scales, by factor of $10^{48}$ for $E_{\rm max}\sim 10^3$~GeV to $10^{124}$ for $E_{\rm max} \sim 10^{19}$~GeV.

It is unlikely that there is technically natural resolution to the cosmological constant's fine tuning problem - this would require new physics at $10^{-3}$~eV. A valiant attempt at such a theory is being made by Sundrum\cite{Sundrum}, but it is highly contrived to have new dynamics at this extremely low scale which modifies only gravity and not the other interactions.

Finally there is a third classic naturalness problem in the Standard Model - that of the strong CP violating parameter $\theta$. It was realized that QCD can violate CP invariance, with a free parameter $\theta$ which can in principle range from zero up to $2\pi$. An experimental manifestation of this CP violating effect would be the existence of a non-zero electric dipole moment for the neutron. The experimental bound on this quantity constrains $\theta \le 10^{-10}$. The quantum corrections to $\theta$ are technically infinite in the Standard Model if we take the cutoff scale $E_{\rm max}$ to infinity. For this reason we would expect that $\theta$ is a free parameter in the model of order unity, and is renormalized in the usual way. However, there is a notable difference with the two other problems above in that if the scale $E_{\rm max}$ is taken to be a very large value the quantum corrections are still quite small. This is because they arise only at a very high order in perturbation theory. So in this case, the quantum correction s do not point to a particular scale at which we expect to find a dynamical solution to the problem.

\section*{Anthropic Constraints}

The standard response to the fine-tuning problems described above is to search for dynamical mechanisms that explain the existence of the fine-tuning. For example, many theories for physics beyond the Standard Model (such as supersymmetry, Technicolor, large extra dimensions etc) are motivated by the desire to solve the fine-tuning of the Higgs vev. These are plausible, but as yet have no experimental verification. The fine-tuning problem for the cosmological constant has been approached less successfully - there are few good suggestions here. The strong CP problem has motivated the theory of axions, in which an extra symmetry removes the strong CP violation, but requires a very light pseudoscalar boson - the axion - which has not yet been found.

However, theories of the multiverse provide a very different resolution of the two greatest fine-tuning problems, that of the Higgs vev and the cosmological constant. This is due to the existence of anthropic constraints on these parameters. Suppose for the moment that life can only arise for a small range of values of these parameters, as will be described below. In a multiverse, the different domains will have different values of these parameters. In some domains, these parameters will fall in the range that allows life. In others, they will fall outside of the range. It is then an obvious constraint that we can only observe those values that fall within the viable range. For the cosmological constant and the Higgs vev, we can argue that the anthropic constraints only allow parameters in a very narrow window, all of which appears to be fine-tuned by the criteria of the previous section. Thus the observed fine-tuning can be thought to be required by anthropic constraints in multiverse theories.

The first application of anthropic constraints to explain fine-tuning was due to Linde and Weinberg\cite{cosmconst}, applied to the cosmological constant even before the this parameter was known to be non-zero. Weinberg in particular gave a physical condition noting that, if the cosmological constant was much different from what it is observed to be,  galaxies could not have formed. The cosmological constant is one of the ingredients that governs the expansion of the universe. If it had been of its natural scale of $(10^3~{\rm GeV})^4$ the universe would have collapsed or blown-apart (depending on the sign) in a small fraction of a second. For the universe to expand slowly enough that galaxies can form, requires $\Lambda$ within roughly an order of magnitude of its observed value. Thus the $10^{124}$ orders of magnitude fine-tuning is spurious - we would only find ourselves in one of the rare domains with a tiny value of the cosmological constant.

Other anthropic constraints can be used to explain the fine-tuning of the Higgs vev. In this case, the physical constraint has to do with the existence of atoms other than hydrogen\cite{agrawal, thibault}. Life requires the complexity that comes from having many different atoms available to build viable organisms. It is remarkable that these atoms do not exist for most values of the Higgs vev, as has been shown by my collaborators and myself. Suppose for the moment that all the parameters of the Standard Model are held fixed, except for $v$ which is allowed to vary. As $v$ gets larger, all of the quark masses grow, and hence the neutron and proton masses also get larger. Likewise the neutron proton mass splitting gets larger in a calculable fashion. The most model-independent constraint on $v$ then comes from the value when the neutron-proton mass splitting becomes larger than the 10~MeV per nucleon that binds the nucleons into nuclei - this comes when $v$ is about 5 times the observed value. When this happens, all bound neutrons will decay to protons\cite{agrawal}. However, a nucleus of only protons is unstable and will fall apart into only hydrogen. The complex nuclei will no longer exist. A tighter constraint takes into account the calculation of the binding energy, which decreases as $v$ increases\cite{thibault}. This is because the nuclear force, especially the central isoscalar force, is highly dependent on pion exchange and as $v$ increases the pion mass also increases, making the force of shorter range and also weaker. In this case, the criteria for the existence of heavy atoms requires $v $ less than about twice its observed value. Finally a third constraint, of comparable strength, comes from the need to have deuterium stable, because deuterium was involved in the formation of the elements in primordial nucleosynthesis and in nucleosynthesis in stars\cite{agrawal,thibault}. In general, even if the other parameters of the Standard Model are not held fixed, the condition is that the weak and the strong interactions must overlap. The masses of quarks and leptons arise in the weak interactions. In order to have complex elements, some of these masses must be lighter than the scale of the strong interactions, and some will be heavier. This is a strong and general constraint on the electroweak scale.  All of these constraints tell us that the viable range for the Higgs vev is not the ~30 orders of magnitude described above, but only the tiny range allowed by anthropic constraints.

\section*{Lack of Anthropic Constraints}

While anthropic constraints have the potential to solve the two greatest fine-tuning problems of the Standard Model, similar ideas very clearly fail to explain the naturalness problem of the strong CP violating parameter $\theta$\cite{dynamics}. For any possible value of $\theta$ in the allowed range from $0\to 2\pi$, there would be little influence on life. The electric dipole moments that would be generated could produce small shifts in atomic energy levels, but would not destabilize any elements. Even if a mild restriction could be found, there would be no logical reason why $\theta$ should be as small as $10^{-10}$. Therefore the idea of a multiverse does nothing to solve this fine-tuning problem.

The lack of an anthropic solution to this problem is a very strong constraint on multiverse theories. It means that in a multiverse ground state that satisfies the other anthropic constraint, the strong CP problem must {\it generically}  be solved by other means. Perhaps the axion option, which appears to us to be an optional addition to the Standard Model, is in fact required to be present for some reason - maybe in order to generate dark matter in the universe. Or perhaps there is a symmetry that initially sets $\theta$ to zero, in which case the quantum corrections shift it only by a small amount. This can be called the ``small infinity'' solution, because while the quantum correction is formally infinite, it is small when any reasonable cutoff is used. Thus the main problem in this solution is to find a reason why the bare value of $\theta$ is zero rather than some number of order unity. In any case, in multiverse theories the strong CP problem appears more serious than the other fine-tuning problems and requires a dynamical solution.

\section*{Physical Mechanisms}

The above discussion can be viewed as a motivation for multiverse theories. Such theories would provide an explanation of two of the greatest puzzles of particle physics. However, this shifts the focus to actually constructing such physical theories. So far we have just presented a ``story'' about a multiverse. It is a far different matter to actually construct a real physical theory that realizes this story.

The reason that it is difficult to construct a multiverse theory is that most theories have a single ground state, or at most a small number. It is the ground state properties that determine the parameters of the theory. For example, the Standard Model has unique ground state, and the value of the Higgs vev in that state determines the overall scale for the quark masses etc. Sometimes theories with symmetries will have a set of discretely different ground states, but generally just a few. The utility of the multiverse to solve the fine tuning problems requires that there be {\it very} many posible ground states. For example if the cosmological constant has a fine-tuning problem of a factor of $10^{50}$, one would expect that one needs of order $10^{50}$ different ground states with different values of the cosmological constant in order to have the likelihood that at least one of these would fall in the anthropically allowed window.

In fact such theories do exist, although they are not the norm. There are two possibilities, ones where the parameters vary continuously and ones where the parameters vary in discrete steps. In the former case the variation of the parameters in space and time must be described by a field. Normally such a field would settle into the lowest energy state possible, but there is a mechanism whereby the expansion of the universe ``freezes'' the value of the field and does not let it relax to its minimum\cite{scalar}. However, since the present expansion of the universe is imperceptibly small, the forces acting on this field must be exceptionally tiny. There is a variant of such a theory which has been applied to the fine-tuning of the cosmological constant. However, it has proven difficult to extend this theory to the the variation of other parameters.

A more promising type of multiverse theory appears to be emerging from string theory. String theory originates as a 10 or 11 dimensional theory, although in the end all but 4 of the spacetime dimensions must be rendered unobservable to us, for example by being of very tiny finite size in the extra dimensions. Most commonly, the extra dimensions are ``compact'', which means that they are of finite extent but with out an endpoint, in the sense that a circle is compact. However, solutions to string theory seem to indicate that there are very many low energy solutions which have different parameters, which depend on the size and shape of the many compact dimensions\cite{landscape}. In fact there are so many that one estimate puts the number of solutions that have the properties of our world, within the experimental error bars for all measured parameters, as of order $10^{100}$. There would then be many more with parameters different from ours outside of the range of our parameters. Truly in this case there are astonishingly many possible sets of parameters for solutions to string theory. This feature of having fantastically many solutions to string theory in which the parameters vary as you move through the space of solutions is colloquially called the ``landscape''.

There are two key properties of these solutions. The first is that they are discretely different, not continuous\cite{polchinski}. The different states are described by different field values in the compact dimensions. These field values are quantized, because they need to return to the same field as one goes around the compact dimension. With enough fields and enough dimensions, the number of solutions rapidly becomes extremely large.

The second key property is that transitions between the different solutions are known\cite{nucleation}. This can occur when some of the fields change their values. From our four-dimensional point of view, what occurs is that a bubble nucleates, in which the interior is one solution and the exterior is the other solution. The rate for such nucleation can be calculated in terms of string theory parameters. In particular, it apparently always occurs during inflation or at finite temperature. Nucleation of bubbles commonly leads to large jumps in the parameters, such as the cosmological constant, and the steps do not always go in the same direction.

These two properties imply that a multiverse is formed in string theory if inflation occurs. There are multiple states with different parameters and transitions between these occur during inflation. The outcome is a universe in which the different regions - the interior of the bubble nucleation regions - have the full range of possible parameters.

String theorists long had the hope that there would be a unique ground state of string theory. It would be indeed wonderful if one could prove that there is only one true ground state and that state leads to the Standard Model with exactly the parameters seen in nature. It would be hard to imagine how a theory with such a high initial symmetry would lead only to a world with parameters with as little symmetry as seen in the Standard Model, such as $m_u = 4$~MeV, $m_d= 7$~MeV, etc.  But if this were in   fact shown it would certainly prove the validity of string theory. Against this hope, the existence of a landscape and a multiverse seems perhaps disappointing. Without a unique ground state, we cannot use the prediction of the parameters as a proof of string theory.

However, there is another sense in which the string theory landscape is a positive development. Some of us who are working ``from the bottom up'' have been led by the observed fine-tuning (in both senses of the word) to desire the existence of a multiverse with exactly the properties of the string theory landscape. From this light, the existence of the landscape is a strong motivation in favor of string theory, more immediate and pressing even than the desire to understand quantum gravity.

Inflation also seems to be a necessary ingredient for a multiverse. This is because we need to push the boundaries between the domains well outside of our observable horizon. Inflation neatly explains why we see a mostly uniform universe even if the greater multiverse has multiple different domains. The exponential growth of the scale factor during inflation makes it reasonable that we see a uniform domain. However, today inflation is the ``simple'' ingredient that we expect really does occur, based on the evidence of the flatness of the universe and the power spectrum of the cosmic microwave background temperature fluctuations. It is the other ingredient, that of having very many ground states of a theory, that is much more difficult.

\section*{Testing through a full theory}

Let us be philosophical for a moment. Anthropic arguments and invocations of the multiverse can sometimes border on being non-scientific. You cannot test for the existence of other domains in the universe outside our visible universe nor can you  find a direct test of the Anthropic Principle. This leads some physicists to reject anthropic and multiverse ideas as being outside of the body of scientific thought. This appears to me to be unfair. Anthropic consequences appear naturally in some physical theories. However, there are nevertheless non-trivial limitations on what can be said in a scientific manner in such theories.

The resolution comes from the realization that neither ``Anthropic'' nor
``Multiverse'' constitute a concrete theory. Instead there are real theories, such as apparently string theory, which have a multiverse property and hence our domain would automatically satisfy anthropic constraints. These are not vague abstractions but are real physical consequences of real physical theories. In this case, ``Anthropic'' and ``Multiverse'' are not themselves the theory but rather the output of a full theory. Our duty as scientists is not to give up because of this but to find other ways to test the original theory. Experiments are reasonably local and we need to find some reasonably local tests that probe the original full theory.

However, it has to be admitted that theories with a multiverse property, such as perhaps the string landscape where apparently ``almost anything goes'', make it difficult to be confident of finding local tests. Perhaps there are some consequences which always emerge from string theory for all states in the landscape. For example, one might hope that the bare strong CP violating $\theta$ angle is always zero in string theory and that it receives only a small finite renormalization. However, other consequences would certainly be of a statistical nature that we are not used to. An example is the present debate as to whether supersymmetry is broken at low energy or high energy in string theory. It is likely that both possibilities are present, but the number of states of one type is likely very different (by factors of perhaps $10^{100}$) from the number of states of the other type - although it is not presently clear which is favored. If this is solved it will be a good statistical prediction of string theory. If we can put together a few of such statistical predictions, we can provide an effective test of the theory.

\section*{Test using quark and lepton masses}

Of the parameters of the Standard Model, none are as confusing as the masses of the quarks and leptons. From the history of the periodic table and atomic/nuclear spectroscopy, we would expect that the masses would show some patterns that reveal the underlying physics. However, no such pattern has ever been found. In this section, I will describe a statistical pattern, namely that the masses appear randomly distributed with respect to a scale invariant weight, and will discuss how this can be the probe of a multiverse theory.

In a multiverse or in the string theory landscape, one would not expect that the quark and lepton masses exhibit any pattern. Rather, they would be representative of one of the many possible states available to the theory. Consider the ensemble of ground states which have the other parameters of the Standard Model held fixed. In this ensemble, the quark and lepton masses are not necessarily uniformly distributed. Rather we could describe their distribution by some some {\it weight}\cite{weight, ddr}. For example perhaps this weight favors quarks and leptons with small masses, as in fact seen experimentally. We would then expect that the quark masses that we see in our domain are not particularly special but are typical of a random distribution with respect to this weight.

The quark masses appear mostly at low energy, yet extend to high energy. To pull out the range of weights that could lead to this distribution involves a detailed study of their statistical properties. Yet it is remarkable easy to see that they are consistent with being scale invariant. A scale invariant weight means that the probability of finding the masses in an interval $dm$ at any mass $m $ scales as $dm/m$. This in turn means that the masses should be randomly distributed when plotted as a function of $\ln(m)$. It is easy to see visually that this is the case. In the Figure, I show the quark and lepton masses plotted on a log scale. One can readily see that this is consistent with being a random distribution in this variable. The case for a scale invariant distribution can be quantified by studying the statistics of 6 or 9 masses distributes with various weights\cite{ddr}. When considering power-law weights, $dm/m^\delta$ one can constrain the power $\delta $ to be greater than 0.8. The scale invariant weight is an excellent fit. Also one may discuss the effects of anthropic constraints on the weights\cite{ddr}.
\begin{figure}[ht]
 \begin{center}
  \includegraphics[scale=0.75]{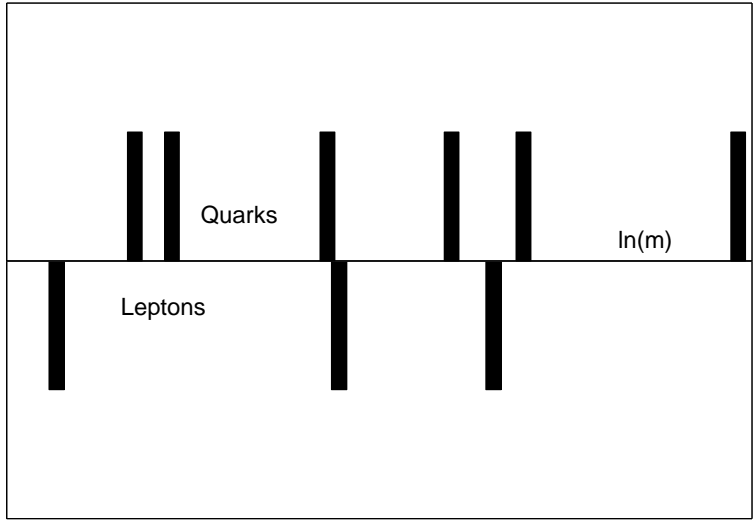}
 \end{center}
 \caption{The quark and lepton masses on a log scale. The result appears
visually to be consistent with a random distribution in $\ln m$, and
quantitative analysis bears this out.}
 \label{scaleinvariant}
\end{figure}

What should we make of this statistical pattern? In a multiverse theory, this pattern is the visible remnant to the underlying ensemble of ground states of different masses. An example of how this distribution could appear from a more fundamental theory is given by the Intersecting Brane Worlds (IBW) solutions of string theory\cite{ibw}. In these solutions, our four-dimensional world appears as the from the intersection of solutions (branes) of higher dimension, much as a one dimensional line can be described as the intersection of two two-dimensional surfaces. In these these theories, the quark and lepton masses are determined by the area between three intersections of these surfaces. In particular, the distribution is proportional to the exponential of this area $m\sim e^{-A}$. In a string landscape there would not be a unique area possible, but rather a distribution of areas. The mathematical connection is that if these areas are distributed uniformly - i.e. with a constant weight - then the masses are distributed with a scale invariant weight. In principle, the distribution of areas is a calculation that could be performed as we understand string theory better. Thus we could relate solutions of string theory to the observed distribution of masses in the real world. This illustrates how we can test the predictions of a multiverse theory without a unique ground state.

\section*{Summary}

The idea of a multiverse can make positive contributions to particle physics. In a multiverse, some of our main puzzles disappear, but they are replaced by new questions.

We have seen how the multiverse can provide a physical reason for some of the fine-tuning that seems to be found in nature.
Note that there are in fact two distinct meanings of the phrase ``fine-tuning'' that are used in different parts of the scientific literature. One meaning, often encountered in discussions of anthropic considerations, relates to the observation that the observed parameters seem to be highly tuned to the narrow window that allows life to exist. The other meaning is the particle physics usage described above concerning the relative size of the quantum corrections compared to the measured value. This latter usage has no a-priori connection to the former. However, the idea of the multiverse unites the two uses - the requirement of life limits the possible range of the particle physics parameters and can explain why the measured values are necessarily so small compared to the quantum effects.

However, in other cases the multiverse makes the problems harder. The strong CP problem is not explained by the multiverse. It is a clue that a dynamical solution to this problem has to be a generic feature of the underlying full theory.

The flavor problem of trying to understand the properties of the quarks and leptons also gets reformulated. I have described how the masses appear to be distributed in a scale-invariant fashion. In a multiverse theory, it is possible that this is a reflection of the dynamics of the underlying theory and that this feature may someday be used as a test of the full theory.

We clearly have more to discover in particle physics. In answering the pressing experimental questions on the the existence of the Higgs boson and the nature of dark matter, etc., we undoubtable will learn more about the underlying theory. We also hope that the new physics that emerges will also shed light on the aesthetic questions concerning the Standard Model. The idea of the multiverse is a possible physical consequence of some theories of physics beyond the Standard Model. It has not been heavily explored in particle physics, yet further presents challenges and opportunities. We clearly have more work to do before we can assess how fruitful this idea will be for the theory of the fundamental interactions.

\section*{Acknowledgement}

I am pleased to thank my collaborators on these topics, Steve Barr, Dave Seckel, Thibault Damour, Andreas Ross and Koushik Dutta, as well as my long term collaborator on more sensible topics, Gene Golowich, for discussions that have helped shape my ideas on this topic. My work has been supported in part by the U.S. National Science Foundation and by the John Templeton Foundation.

\end{document}